\documentclass[aps,pre,groupedaddress,amssymb,amsmath,pra,floatfix,showpacs]{revtex4}
\usepackage{epsfig}
\usepackage{xcolor}

\begin{document}

\title{Motion of a superconducting loop in an inhomogeneous magnetic field:\\
 a didactic experiment}
\author{Marco Giliberti, Luca Perotti$^{a}$, and Lucio Rossi}
\affiliation{Dipartimento di Fisica dell'Università di Milano, via Celoria 16, 20133 Milano Italy.}
\affiliation{$^{a}$Department of Physics, Texas Southern University, Houston,Texas 77004 USA.}
\date{\today}

\begin{abstract}
We present an experiment conductive to an understanding of both
Faraday's law and the properties of the superconducting state. It
consists in the analysis of the motion of a superconducting loop moving under the influence of gravity in an inhomogeneous horizontal magnetic
field. Gravity, conservation of magnetic
flux, and friction combine to give damped harmonic oscillations.
The measured frequency of oscillation and the damping constant as a function
of the magnetic field strength (the only free parameter) are in good
agreement with the theoretical model. 
\end{abstract}


\maketitle

\section{Introduction}

High temperature superconductors have not only brought new applications
and an increase of the general interest in superconductivity, but
also the possibility to introduce with relative ease university students
to actual experiments. 

A number of didactic experiments have been described in the literature
\cite{esperimenti} and laboratory kits are available for many of
them \cite{kits}.

The following experiment can help students get acquainted with superconductivity
by describing and visualizing in a fascinating way one of the most
well known aspects of superconductivity: magnetic levitation induced
by super-currents. While conceptually simple, the complexity of the
apparatus --even in its simpler high temperature version-- makes it
suitable for a graduate lab; to make it more accessible, video recordings
can be made that could be used for in class presentations.

Due to Faraday's law, a time dependent magnetic field induces currents
in a conducting circuit. The same result is obtained when the circuit
moves in an inhomogeneous magnetic field. In the case of an ohmic
conductor, these currents are heavily damped by the circuit resistance
$R$, as the damping time constant L/R --$L$ being the circuit inductance--
is usually very small. However in the case of a superconducting loop
the situation is different. In fact in this case --as the resistance
is vanishingly small-- the induced currents are persistent and the
magnetic flux through the loop is therefore constant.

Allowing a conducting loop to fall under the influence of gravity
in a decreasing inhomogeneous magnetic field, the induced current
causes a force on the loop opposing its downwards acceleration. In
the case the loop is superconducting, the only damping is mechanical
and for suitably chosen parameters such that the magnetic force is strong enough
to reverse the acceleration of the loop, the loop itself hangs unsupported
while oscillating around its equilibrium point for sufficiently long
times to allow detection of the oscillations. Note that this
is a very different kind of ``levitation" from the
one seen in the didactic levitation experiments usually found in the literature \cite{esperimenti},
which is due to the Meissner effect \cite{meissner} (expulsion
of the magnetic field from the superconductor).

Starting from an idea of D.McAllister \cite{mca} and R. Romer \cite{rom}
we performed a real experiment where a superconducting loop falls
in a magnetic field with a sharp discontinuity.

\section{Theoretical model}

\label{due}

In our experiment the loop moved in liquid helium (LHe) and not in
vacuum, as discussed by Romer \cite{rom}, and thus his model must
be modified to take into account the dissipation due to the viscosity
of the medium and the friction between the loop and parts of the apparatus.
\textit{Using a high temperature conductor permits to avoid the use
of LHe; in the following sections we shall indicate in italics changes
in the apparatus and procedure for this simpler experiment.}

\begin{figure}[htbp]
\centering\epsfig{file=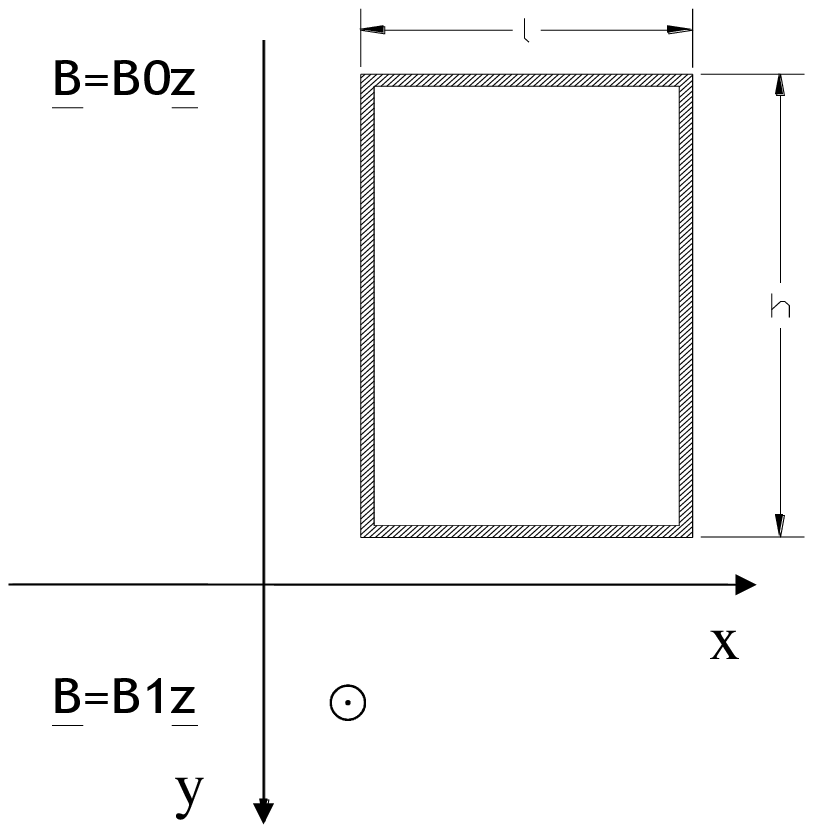,width=0.7\linewidth} \caption{The superconducting loop of height $h$ and width $l$, and the reference frame of the model.}
\label{fig1} 
\end{figure}

Let's consider a rectangular loop of height $h$ and width $w$ that
moves in a given medium under the influence of gravity (directed downward)
in a magnetic field $\bar{B}$ (directed out of the page) as in Figure
\ref{fig1}.

Let's moreover assume that:

1. the magnetic field $\bar{B}$ is horizontal, uniform and of constant
magnitude $B_{0}$ for $z\leq0$ and $B_{1}<B_{0}$ for $z>0$ (note
that the $z$ axis is pointing downward);

2. The loop is kept vertical and its motion is along the vertical
axis so that the magnetic force is also vertical and the problem can
be considered to be one dimensional;

3. for simplicity, all damping effects, even those due to friction
with the apparatus, are described as a single viscous force $F_{d}=-\eta\dot{z}$
linear in the loop vertical velocity $\dot{z}$ where the dot denotes,
as usual, differentiation with respect to time.

Suppose the loop is released from rest with its lower edge in $z=0$
with no initial current. As it falls down, it progressively leaves
the region with magnetic field $B_{0}$ and enters the zone with field
$B_{1}$ and a current $I$ begins to circulate, in order to maintain
constant the initial magnetic flux. When the lower edge of the loop
is in $z\in[0,h]$, the concatenated outward flux of $B$ is given
by: 
\begin{equation}
\phi=B_{0}lh+(B_{1}-B_{0})lz+LI,\label{one}
\end{equation}
where $I$ is the current (taken positive if flowing counterclockwise)
and $L$ is the self-inductance of the loop. Therefore, denoting by
$m$ the mass and by $R$ the resistance of the loop, from Faraday's
law we have: 
\begin{equation}
RI=-\dot{\phi}=(B_{0}-B_{1})l{\dot{z}}-L{\dot{I}}.\label{two}
\end{equation}
Note that when the loop leaves the region $z\in[0,h]$, the current
does not depend on $z$ any longer but becomes constant: $I=0$ for
$z<0$ and $I=(B_{0}-B_{1})lh/L$ for $z>h$. Care must therefore
be taken to design the experiment to avoid these discontinuities and
keep the motion of the loop within $z=0$ and $z=h$. We now write
Newton's law of motion for the loop, subjected to the three forces
gravitational, magnetic, and viscous: 
\begin{equation}
m{\ddot{z}}=mg-\eta{\dot{z}}-(B_{0}-B_{1})lI,\label{three}
\end{equation}
where $g$ is the usual gravitational acceleration on the earth surface.
Differentiating equation (\ref{two}) and using equations (\ref{two})
and (\ref{three}) to eliminate ${\dot{z}}$, we obtain for $I$ the
well known equation of damped harmonic oscillations: 
\begin{equation}
{\ddot{I}}+2\gamma{\dot{I}}+\omega^{2}I=F.\label{four}
\end{equation}
with: 
\begin{eqnarray*}
\gamma\equiv\frac{1}{2}\left({\frac{R}{L}+\frac{\eta}{m}}\right);\\
\omega\equiv\left[{\frac{\eta R}{mL}+\frac{(B_{0}-B_{1})^{2}l^{2}}{mL}}\right]^{1/2};\\
F\equiv\frac{gl(B_{0}-B_{1})}{L}.
\end{eqnarray*}

The equation for $z$ is in general more complicated than above, depending on
a time-dependent driving force. Luckily superconductivity simplifies
things: when $R=0$, equation \ref{two} reduces to the proportionality
relation $\dot{z}=\dot{I}L/[l(B_{0}-B_{1})]$. Since we have chosen
both initial conditions $z_{0}$ and $I_{0}$ to be zero, the same
proportionality relation holds between $z$ and $I$, and the resulting
equation of motion is again the equation for damped harmonic motion:
\begin{equation}
{\ddot{z}}+2\gamma{\dot{z}}+\omega^{2}z=g,\label{five}
\end{equation}
the only difference with eq. \ref{four} being in the driving term
that is now just $g$ itself. If we now suppose that the damping is
less than critical, i.e. if $\gamma^{2}<\omega^{2}$, equation \ref{five}
has the following solution: 
\begin{equation}
z(t)=\frac{g}{\omega^{2}}\left[{1-e^{-\gamma t}\left({\cos{\Omega t}+\frac{\gamma}{\Omega}\sin{\Omega t}}\right)}\right]=\frac{g}{\omega^{2}}\left[{1-\frac{e^{-\gamma t}}{\cos\varphi}\cos{(\Omega t-\varphi)}}\right],\label{six}
\end{equation}
where 
\begin{eqnarray}
\Omega=(\omega^{2}-\gamma^{2})^{1/2},\label{seven}\\
\varphi=\arctan\frac{\gamma}{\Omega}.
\end{eqnarray}
If this is this case, as stated in the introduction, the loop performs
oscillations with a frequency $\Omega$ and damping time constant
$\tau=\gamma^{-1}$ which is almost entirely due to mechanical friction
(some tiny electrical dissipation is present when using a type II
superconductor, such as Niobium).

Note that the initial amplitude of the oscillations is 
\begin{equation}
\frac{g}{\omega^{2}}=\frac{gmL}{(B_{0}-B_{1})^{2}l^{2}}
\end{equation}
which means that to keep the loop in the region of validity of eq.
\ref{six}, we need 
\begin{equation}
(B_{0}-B_{1})>\frac{1}{l}\sqrt{\frac{gmL}{h}}.\label{B}
\end{equation}

\section{Experimental set up}

\label{tre}

\begin{figure}[htbp]
\centering\epsfig{file=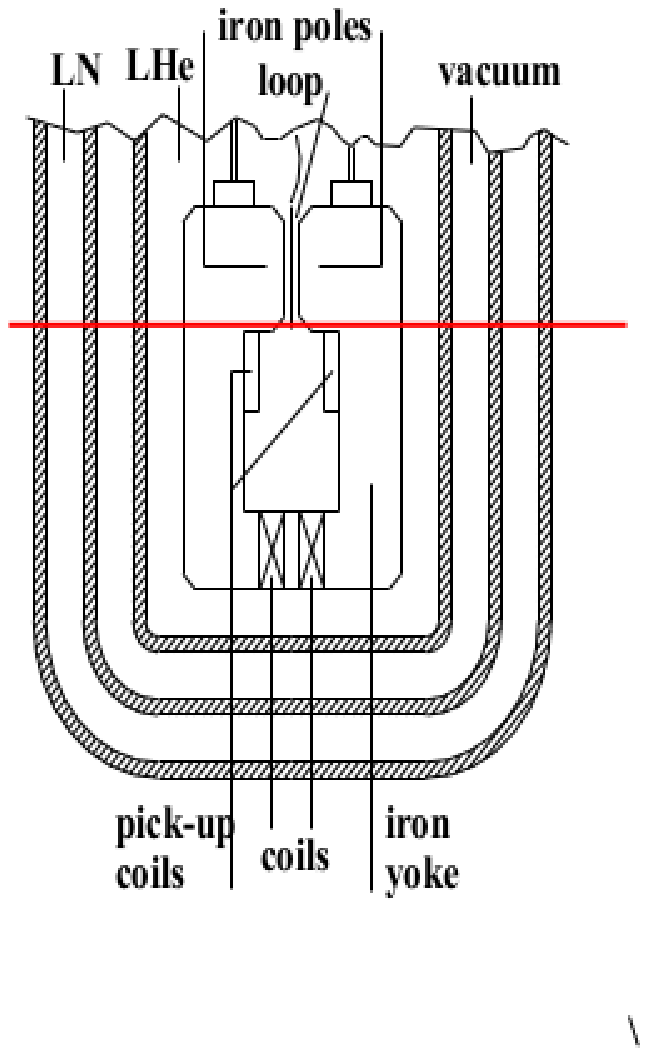,width=0.5\linewidth} \caption{Schematics of the bottom part of the apparatus. The double walls between
the LN chamber and the LHe one are shown while the outer double walls
are not. The horizontal red line marks the transition from the $B_0$ (above) to the $B_1$ magnetic field (below).}
\label{fig2} 
\end{figure}

The experiment has been performed with a superconducting loop cut
out of a niobium (Nb) sheet weakly alloyed with 1\% titanium, immersed
in a LHe bath at $4.2 K$, well below the Nb critical temperature (about
$9 K$). In order to minimize the cryogenic losses of LHe, we used a
cryostat where the helium vessel was surrounded by a vessel filed
with liquid nitrogen (LN); both vessels were moreover double wall
ones, with a vacuum chamber as insulation. A schematic view is shown
in Figure \ref{fig2}.

{\it A HTS (High Temperature Superconductor) material could be used for
the superconducting loop. The most suitable candidate is the YBCO
(Yttrium based copper oxide) coated conductor, since it is produced
in thin sheets so it is relatively easier to obtain a small loop without
resistive joints. YBCO has a critical temperature of $85 K$, therefore
the experiment could be done directly in liquid nitrogen, highly simplifying
the cryostat and also reducing the cost of the experiment. Using an
YBCO loop may enable even to design the experiment without cryogenic
fluid: the loop could be placed in an inner chamber without cryogen
and cooled by the LN surrounding the chamber via radiation and conduction
(the loop could be in a N or He atmosphere). As a further advantage,
in this configuration the friction would be greatly reduced (no liquid
in the experiment chamber, only gas). The cooling of the chamber via a cryocooler, beside eliminating the need of cryogenic fluid, greatly reduce the complexity and operational risk. Use of a
cryocooler would facilitate the carrying out of the experiment at
variable temperature. In this case it would also be easy to show what
happens when the temperature is slowly increased above the critical
temperature. }

In the inner zone of the cryostat we placed an electromagnet, which
generated the needed magnetic field. The poles of the magnet were
$6 mm$ apart and the cross-section of the gap was $30\times40 mm^2$.
To measure the magnetic field in operation condition, a cryogenic
Hall probe was placed in a sample holder that could be moved --acting
from outside the cryostat-- along the axis of the magnet to map the
field itself. The superconducting loop, cut from a Nb sheet, had the
following dimensions: $w=25 mm$, $h=39 mm$. Its mass was $m=0.93 g$
and its self-inductance $L$ was $6.4\pm0.5\times10^{-8} H$; the minimum
value of $(B_{0}-B_{1})$ required to keep the loop in the operation
region is therefore about $5 mT$. Two guides kept the loop away from
the edges of the magnet poles and limited unwanted lateral movements.
They were made of G10 fiberglass and, to reduce friction, they were
covered with Molykote (bi-solphure of molybdenus with some graphite
powder).

To reduce the boiling of LHe in the zone of the moving loop (that
could disturb the oscillations and their visibility), the two coils
of the magnet have been wound using very low resistivity wires so
as to give the minimum heat production once the magnetic field, suitable
to keep frequency of oscillations in a ``reasonable"
range, has been turned on. For $B_{0}\sim15 mT$, the calculated evaporation,
due to the heat production of the coils, was about $3\times10^{-6} m^3/h$.

Care has also been taken of always being below the critical magnetic
field (which for Nb is about $180 mT$ with a density of current of
$1500 A/mm^{2}$). The ``reasonable" frequency of oscillation
has been chosen to be about $10 Hz$, so that the motion could be video-recorded,
both for didactic purposes \textbf{(e.g. in--class presentations)}
and as a set up and measuring tool, as we shall explain in the following.
For this purpose, a common $75 frames/s$ video recorder was used.

The cryostat was designed with a double optical window, through the
LN vessel and the LHe chamber, to allow a direct observation of the
oscillating loop in the magnet gap between poles. The video camera
could be placed in front of it, as a visual aid when setting up the
experiment and to record the oscillations of the loop.

The motion of the loop was detected measuring the voltage induced
in two pick-up coils by the magnetic flux that was generated by the
supercurrents in the moving loop. These coils consisted of $1000$
windings of \textbf{$0.2 mm$} diameter copper wire surrounding an
area of $25.8 \times11.1 mm^2$, for a total thickness of $3.5 mm$.
They were placed just below the magnet poles, where the motion of
the loop took place. The voltage in the coils was registered by an
oscilloscope (National VP-5730A ID4N0073B12). As the current in the
pick-up coils was neglible, so were the heating around them and the
braking effect due to the magnetic field they generated.

\section{Experimental procedure}

\label{qua}

The experiment can be divided into four time--ordered steps.

1. The loop was placed inside the magnet gap and kept in position
by a thin nylon thread fastened to the top of the cryostat.

2. The electromagnet was switched on at a given magnetic field.

3. Vacuum was made in the vacuum chamber; the external section
of the cryostat was then filled with liquid nitrogen; 
finally, LHe was poured into the inner zone so that the transition
of the loop to the superconducting state could take place.

Let us notice that steps $1$ and $2$ had to precede step $3$ because
we had to place the loop in position, and in the desired magnetic
field, while it was still warm. The reason why the loop had to become
superconducting when the magnetic field was already linked to the
loop itself is that inserting an already superconducting loop in a
field region would have induced a super-current tending to expel the
loop itself. 

4. The thread, used to hang the loop in the proper position, was suddenly
disconnected from the top of the cryostat, so that the loop was let
free to fall down. As predicted the loop started to oscillate before
stopping at the equilibrium position.

\begin{figure}[htbp]
\centering\epsfig{file=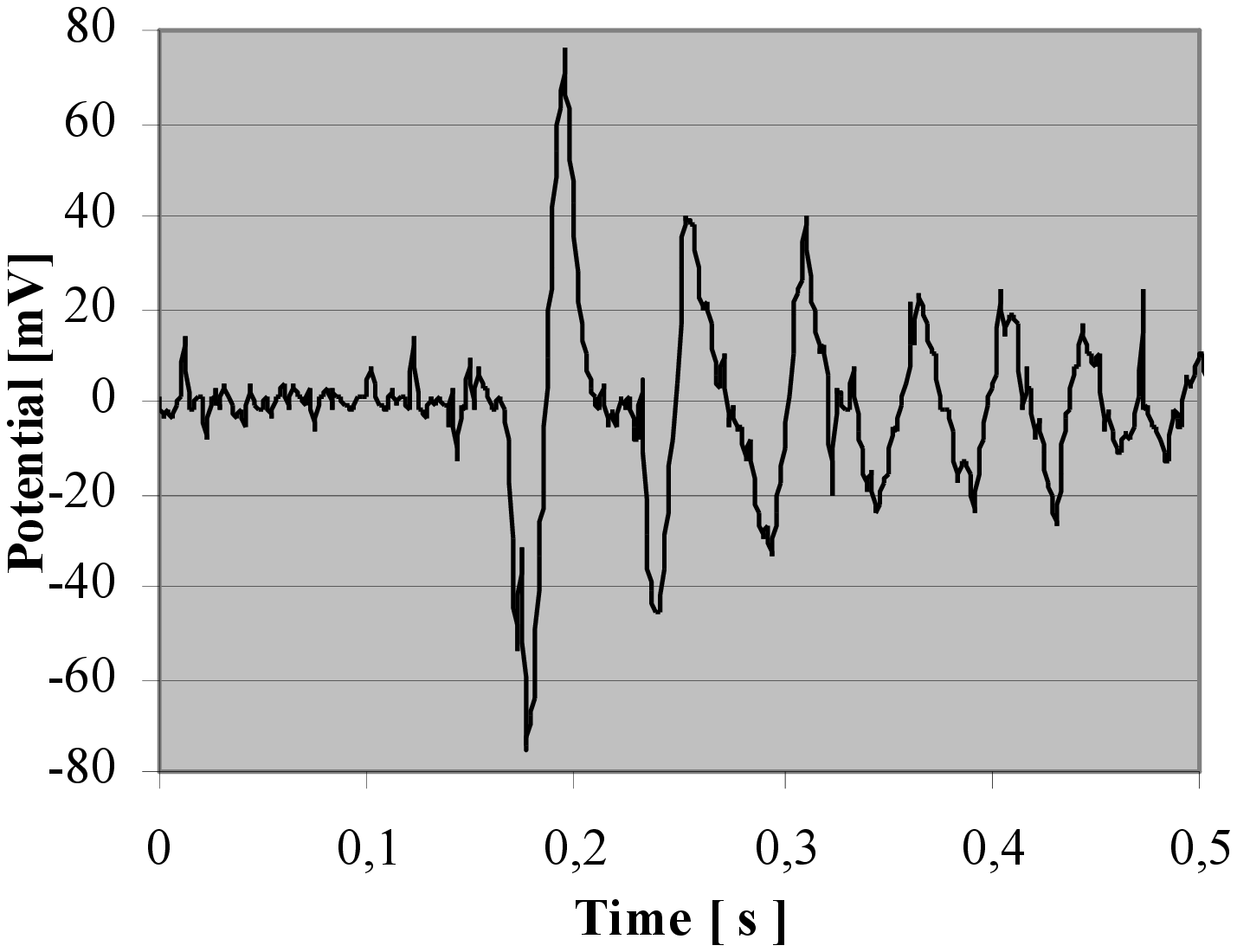,width=1.0\linewidth} \caption{A plot of oscillation as detected by the pick up coils at $B_{0}-B_{1}=35.84\pm 0.75 mT$.}
\label{fig3} 
\end{figure}

The use of the video camera was fundamental since it helped to control
the filling of the cryostat and the positioning of the loop during
the setup, and to afterwards check the motion of the loop itself during
the experiment. Moreover, looking at the motion frame after frame,
we could also get some rough measurements of the frequencies of oscillation
of the loop. Of course, more refined measurements were made by means
of the pick-up coils, connected to the digital oscilloscope, as described
in Section \ref{tre}. The plots generated by the oscilloscope (see
Figure \ref{fig3}) showed the typical behavior of damped harmonic
motions: the time between two maxima was the period of the oscillation
while the damping constant could be evaluated from the decreasing
amplitude of the peaks.

Unfortunately the video camera and the pick-up coils could not be
used at the same time, since the light source, essential to illuminate
the inner cryostat, caused too much noise in the signal coming from
the pick-up coils. Therefore, for each value of the magnetic field,
the experiment had to be divided into two parts: the first one
used the video camera for a first analysis of the motion; the
second one, without the camera but with the pick-up coils switched on,
measured the frequencies from the pick-up coils signal.

\section{Data analysis and results}

\label{cin}

We now discuss the results of the experiment, comparing the measured
oscillation frequencies with the ones predicted by our model.
To determine these latter values, the quantity $B_{0}-B_{1}$ is crucial.
For our purposes the magnetic field $B_{0}$ could be considered uniform
inside the magnet: for a typical operating value $B_{0}=30 mT$, $B_{0}$
is uniform up to $4\%$ all through the magnet gap.

The situation was different outside the magnet poles, where the magnetic
field varied with the position up to $40\%$. Since our model assumes
a magnetic field of uniform strength $B_{1}$ outside the region between
the magnet poles, a suitable average had to be considered. Our choice
was to perform a spatial average limited to the region accessed by
the loop. The vertical range of motion outside the magnet gap has
been visually assessed with the help of the video camera to be $30 mm$.
For this and similar measurements, a graduated scale, calibrated to
match a scale placed in the cryostat before filling it and then removed,
was attached to the camera monitor.

The oscillation frequencieswere measured for $11$ values of
$B_{0}-B_{1}$, varying from $14.8 mT$ to $44.2 mT$. For each value of the
field, $10$ to $20$ tests were conducted. Measurements taken in
the range $5.0 mT <B_{0}-B_{1}<14.0 mT$ were dropped for two reasons;
the first one being that $B_{0}-B_{1}$ was too close to the $5.0
mT$ lower bound to keep the loop in the operating region. The second
reason is that the records taken by the video-camera show that at
such low values of $B_{0}-B_{1}$ the loop often went partially off
the guides and oscillated on a non-vertical plane, contrary to the
assumptions of the theoretical model we used.

\begin{figure}[htbp]
\centering\epsfig{file=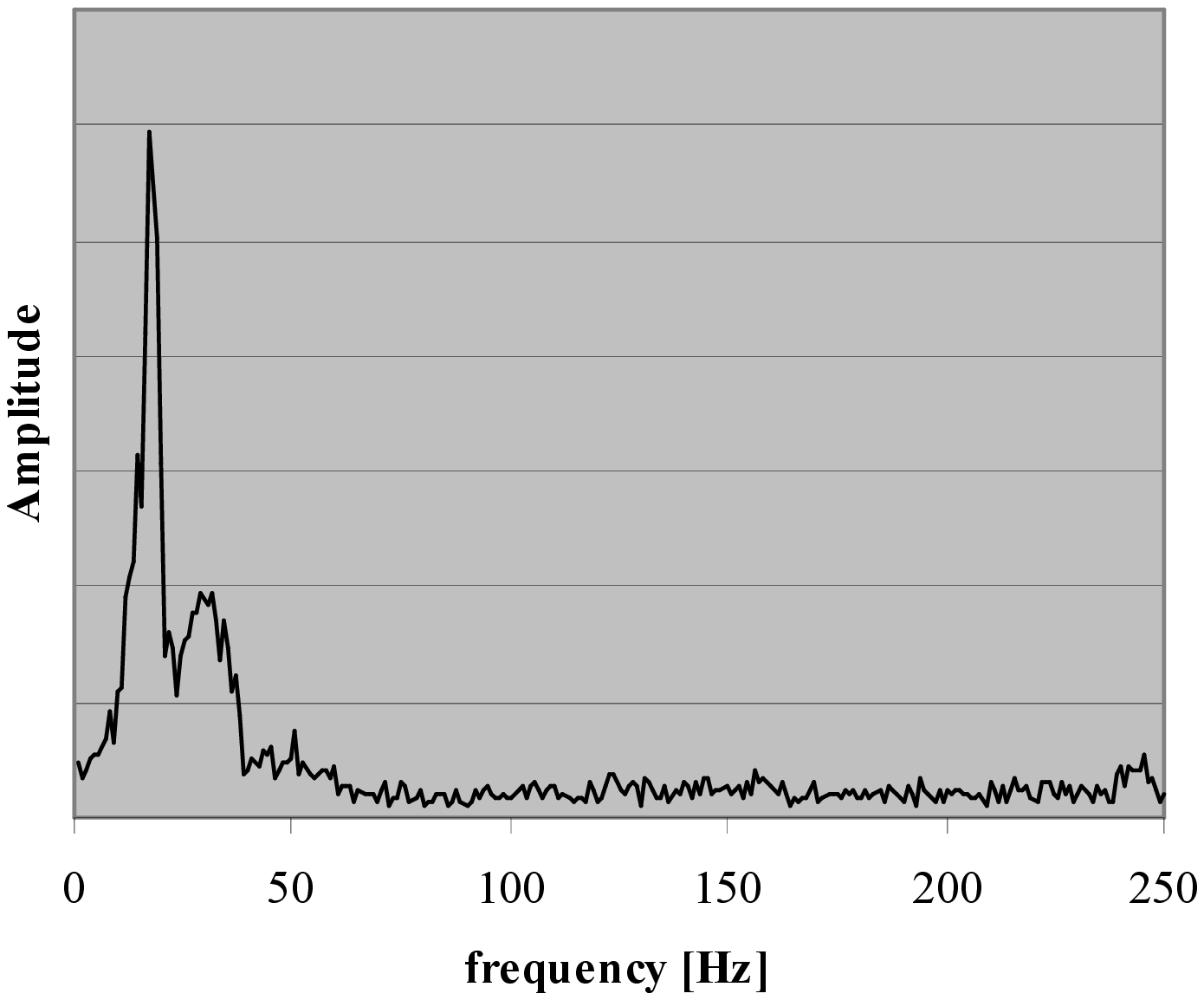,width=0.8\linewidth} \caption{Oscillation spectrum at $B_{0}-B_{1}=35.84\pm0.75 mT$.}
\label{fig4} 
\end{figure}

Two analysis methods were used to get the oscillation frequency,
both of them from the pick-up coils voltage measurements: spectral analysis,
using the \textit{Microsoft Excel $\copyright$} Fourier analysis
instrument (Figure \ref{fig4} shows a typical example of the spectra
we obtained) and the direct measurement on the plot of the oscillation period
(``plot method"). The ``plot method"
has been mainly used as control of the numerical procedure.

The calculated spectra showed various peaks of different amplitude
but only the first two in each plot were clearly stronger than the others: see
Figure \ref{fig4}. We consistently chose the frequency of the lowest frequency
peak as the oscillation frequency of the loop, even in the  two cases ($B_{0}-B_{1}=19.7
mT$ and $B_{0}-B_{1}=26.0 mT$) when the two strongest peaks had nearly the same amplitude.
 The second peak does not usually appear to be harmonic of the first one; we therefore surmise it might be due to the non-uniformity of the $B_1$ field.

The experimental damping time--constant $\gamma$ was again obtained
in two ways: through the ``plot method", i.e. measuring,
from the plots of the pick-up coils voltage, the time it takes for
the oscillations to decay to half their initial amplitude, and --when
the shape of the main spectral peak was regular enough-- through a
measure of its half width. As in the previous case, the second method
was used for control of the procedure.

\quad

\begin{figure}[htbp]
\centering\epsfig{file=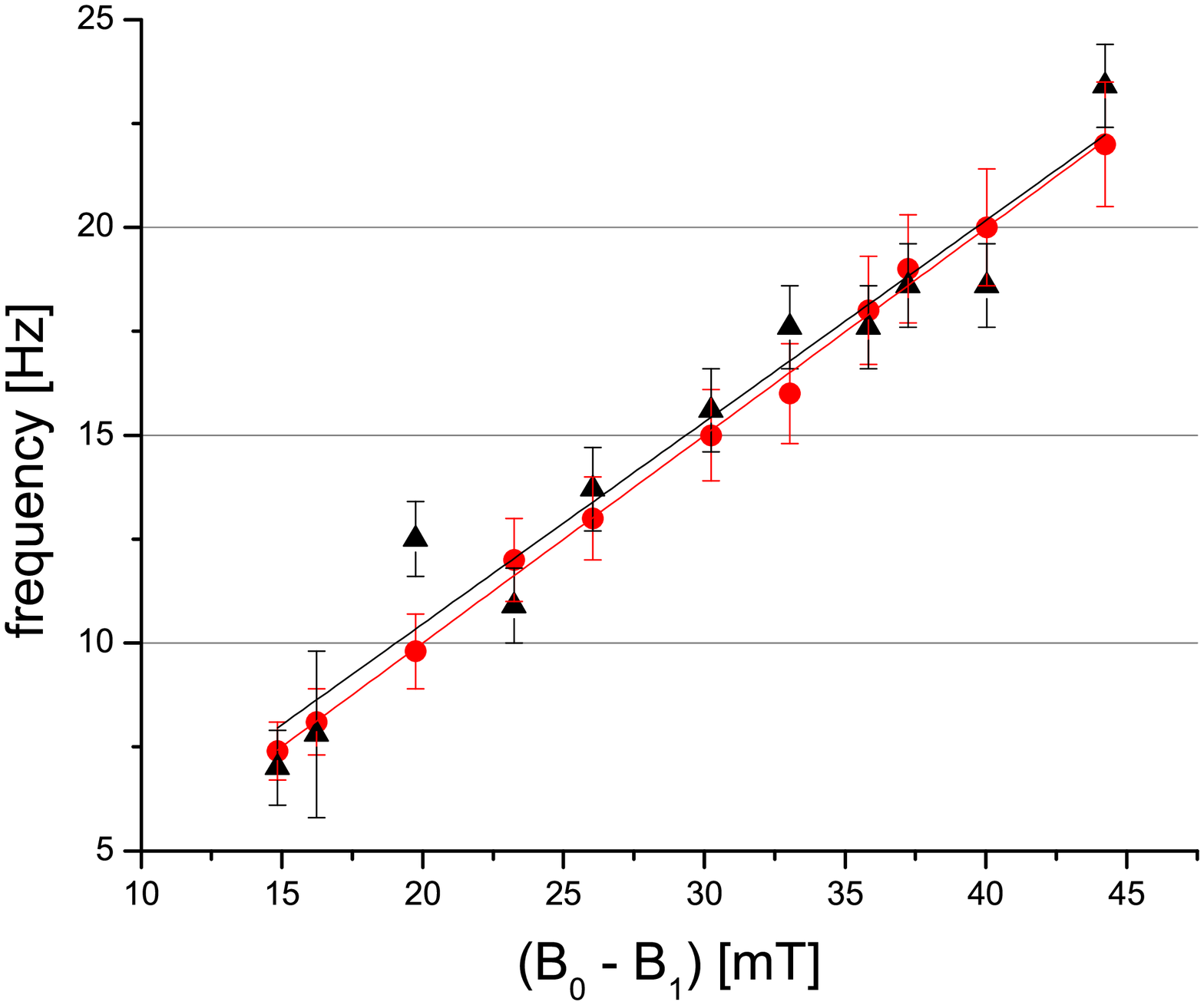,width=1.0\linewidth} \caption{Plot of the loop oscillation frequency vs. magnetic field difference
$(B_{0}-B_{1})$. The black triangles represent direct experimental data,
while the red dots represent the ``theoretical results".
The two lines represent the least square fit of each data set.}
\label{fig5} 
\end{figure}

Our results are summarized in Table \ref{table1} and Figure \ref{fig5},
where we compare our experimental frequencies to the ``theoretical"
frequencies, evaluated from equation \ref{seven}, with $B_{1}$ calculated
as explained above and using the experimental values of $\gamma.$
The listed error for the ``theoretical" frequencies
is due to the uncertainties in these two quantities. In particular,
for low magnetic fields it is mainly due to the error in the
measurements of \textbf{$\gamma$, }while for large magnetic fields
it is mainly due to the uncertainty in the value of\textbf{ $B_{1}$.}

The agreement between the two sets of data in Figure \ref{fig5} is
almost everywhere good; the least square fitting for the experimental
data gives a slope of $0.485\pm 0.041 Hz/mT$ ($R^{2}=0.934$) in
good agreement with that obtained from the fitting of the ``theoretical"
data: $0.00499\pm 0.0075 Hz/mT$ ($R^{2}=0.998$).

\begin{table}
\begin{tabular}{cccccc}
Magnet  & $B_{0}-B_{1}$  & Theoretical  & \hspace{0.05in}  & Spectrum  &  analysis \tabularnewline
current $[A]$  & $[\pm0.75 mT]$  & model  & \hspace{0.05in}  & \qquad First  & peak \quad \quad\tabularnewline
\hline &  &  &  &  &  \tabularnewline
\hspace{0.05in}  & \hspace{0.05in}  & $\nu[Hz]$  & $\sigma_{\nu}[Hz]$  & $\nu[Hz]$  & $\sigma_{\nu}[Hz]$  \tabularnewline
\hline\hline &  &  &  &  &  \tabularnewline
0.9  & 14.84  & 7.4  & 0.7  & 7.0  & 0.9\tabularnewline
1.0  & 16.24  & 8.1  & 0.8  & 7.8  & 2.0\tabularnewline
1.25  & 19.74  & 9.8  & 0.9  & 12.5  & 0.9\tabularnewline
1.5  & 23.24  & 12  & 1.0  & 10.9  & 0.9\tabularnewline
1.7  & 26.04  & 13  & 1.0  & 13.7  & 1.0\tabularnewline
2.0  & 30.24  & 15  & 1.1  & 15.6  & 1.0\tabularnewline
2.2  & 33.04  & 16  & 1.2  & 17.6  & 1.0\tabularnewline
2.4  & 35.84  & 18  & 1.3  & 17.6  & 1.0\tabularnewline
2.5  & 37.24  & 19  & 1.3  & 18.6  & 1.0\tabularnewline
2.7  & 40.04  & 20  & 1.4  & 18.6  & 1.0\tabularnewline
3.0  & 44.24  & 22  & 1.5  & 23.4  & 1.0\tabularnewline
\end{tabular}
\caption{Summary of the experimental results.}
\label{table1} 
\end{table}

\section{Conclusions}

\label{sette}

Due to easy to perform experiments and the availability of cheap laboratory kits, superconductivity has recently become part of undergraduate lab courses and courses for future physics teachers. Here, we have described a didactic experiment that has the advantage of combining superconductivity with another common lab topic --damped harmonic oscillations-- thus giving students the opportunity to deepen their understanding of both. 
Besides giving the students the chance to tackle many different physics topics such as magnetism, superconductivity, electric currents and oscillations, the experiment we propose involves different detecting devices and data analysis techniques that could easily be presented to students in an almost peer to peer instruction mode.

%

The experimental data we collected show that the theoretical model we adopted is sufficiently good and
that it is therefore suitable for a discussion with college
students on superconductivity, Faraday's law and oscillatory motion based on it and on our experimental set-up.
Clearly we could try to improve our model and the fitting of the data
but, for that purpose, more refined mathematics should be used and
it is our opinion that it could obscure more than lighten the physics
involved. 

Following the present technical analysis, we are planning to soon report on 
the in-class experimentation that we are organizing and that will help us decide whether our experiment can help students to get closer to the fascinating phenomena of superconductivity through an ``unusual" application.


\section{Acknowledgments}

\label{nove}

We are very grateful to Mr. Castellazzi and Mr. Ormenese of Videotime
s.p.a. for their precious help and also to the technicians of L.A.S.A.
laboratory for their work and kind suggestions. We also thank Mr.
Giuseppe Baccaglioni of L.A.S.A.for having built the pick-up coils
and measured the self inductance of the loop.

\end{document}